# Multiscale Simulation of Quantum Nanosystems: Plasmonics of Silver Particles

D. Balamurugan and Peter. J. Ortoleva*

Department of Chemistry, Indiana University, Bloomington, IN 47405

*email:ortoleva@indiana.edu

**ABSTRACT**

Quantum nanosystems involve the coupled dynamics of fermions or bosons across multiple scales in space and time. Examples include quantum dots, superconducting or magnetic nanoparticles, molecular wires, and graphene nanoribbons. The number ($10^3$ to $10^9$) of electrons in assemblies of interest here presents a challenge for traditional quantum computations. However, results from deductive multiscale analysis yield coarse-grained wave equation that capture the longer-scale quantum dynamics of these systems; a companion short-scale equation is also developed that allows for the construction of effective masses and interactions involved in the coarse-grained wave equation. The theory suggest an efficient algorithm for simulating quantum nanosystem which is implemented here. A variational Monte Carlo method is used to simulate the co-evolution of long- and short-scale processes. The approach does not require experimental data for calibration. It is validated via experimental data and TDDFT predictions on the nanoparticle size dependence of the plasmon spectrum.

# I. INTRODUCTION

Quantum nanosystems are materials of contemporary interest. Examples are quantum dots[1, 2], superconducting[3] and magnetic[4] nanoparticles, graphene nanoribbons[5], and molecular wires[6]. These systems typically involve $10^3$ to $10^9$ electrons. Their behavior follows from the interplay of collective and dressed particle behaviors. These behaviors are not readily simulated via conventional quantum computations due to the number of electrons and space-time scales involved. Recent studies show the potential for efficient quantum nanosystem computation via deductive multiscale analysis (DMA) and implied computational algorithms[7-11].

DMA yields a coarse-grained wave equation (CGWE) and algorithms for constructing all factors in them from those in the underlying Schrödinger equation [9]. Constructing these factors involves the use of solutions to a short-scale wave equation. The objective of this study is to implement this theory as a multiscale quantum simulation algorithm and test it with data on plasmons in silver nanoparticles.

Multiphysics approaches such as QM/QM[12-14] involve semi phenomenological coupling of models from different scales. For example, to model a local site of reaction in a larger system, a high accuracy quantum calculation is used to characterize a local site of interest, which is then joined phenomenologically to a lower accuracy solution in the remainder of the system. In contrast, the present multiscale approach avoids such phenomenological coupling by deriving and then solving long- and short-scale equations in all regions of the system. The approach for integrating the short- and long-scale information follows from a multiscale perturbation analysis that starts with the many-electron wave equation[9].

The size of nanoparticles observed to express plasmonic behaviors range from 2 to 200 nm[15-18]. In this range, an approach based on the Jellium model[19, 20] captures much of the plasmonic behavior of metal nanoparticles. This model replaces the ions by a uniform charge background, and the electronic states are solved via time dependent density functional theory. The theory is phenomenological in the sense that the ionic-background potential of a particular metal is tuned to match the experimental work function[19]. The predictive power of such an approach is limited due to the need of experimental work function data for calibration. The influence of surface roughness or internal non-uniformity on plasmon behavior is also difficult to address.

Phenomenological electromagnetic theory approaches are used to simulate plasmons in nanostructures[21]. Such approaches include Mie scattering theory[22-24], discrete dipole approximation (DDA)[25-27], and finite difference time domain methods (FDTD)[28, 29]. These methods require experimental data to simulate the electromagnetic response of the system. Mie scattering theory provides analytical solution for the electromagnetic wave scattering by spherical and cylindrical particles which are homogenous or of core-shell structure. The scattering cross section is computed for a given material whose complex refractive index at a particular wavelength is known from experiments. DDA computes the scattering and absorption of electromagnetic waves by nanostructures of arbitrary geometry and composition given the experimental complex refractive index. The nanostructure is represented by a set of dipoles. The discrete dipoles interact with each other and with an imposed electromagnetic field. The system is then described by a set of linear equations to be solved to obtain the polarization. The latter yields scattering amplitude and extinction coefficients for a single nanoparticle or an assembly thereof. In the FDTD approach, Maxwell's equations are solved in the time domain on a rectangular finite difference grid. The electric and magnetic fields are evaluated on each point of the grid and propagated forward in time. The material is specified via the frequency-dependent dielectric function derived from experimental data. Since the spatial region is discretized in the FDTD approach, nanostructures of arbitrary shape can be investigated.

The above quantum and classical methods predict plasmon behaviors in nanoparticles but suffer from one or more of the following limitations. (1) As plasmons are excitations of the electronic system, they should ultimately be described as solutions to the Schrödinger equation. (2) Experimental data is required to calibrate the models for each material and each geometry. (3) Lack of atomic detail makes it difficult to capture the effects of surface roughness, coatings, and internal compositional inhomogeneity.

The theory implied by DMA is fundamentally multiscale. Thus, it yields both the coarse-grained quantum state and the finer-scale one, and captures the transfer of information between scales. The theory is "calibration-free" in the sense that only those factors in the many-electron Schrödinger equation are needed. However, simplification of the multiscale theory can be accomplished using pseudopotential approximations; these are obtained via calibration with more complete quantum calculations or experimental data, and are available for a variety of materials[30]. For example, it is shown here that the properties of plasmons in silver nanoparticles can be predicted using only published data on the ion core pseudopotential. Although the present study focuses on homogeneous silver nanoparticles, the theory and computational algorithm are applicable to complex systems including systems with internal compositional variation, interfaces, defects, and surface roughness.

The DMA approach is reviewed (Sect. II). The related computational algorithm is outlined including the construction of effective masses and interactions (Sect. III). The latter is demonstrated for plasmons in silver nanoparticles (Sect. IV). Conclusions are drawn in Sect. V.

## II. DEDUCTIVE MULTISCALE ANALYSIS

Recently, a method was developed to transform the Schrödinger equation into a form that reveals the multiple scale character of the wavefunction, and enables one to construct its dependencies quantitatively [9]. This deductive approach is briefly described below.

The $N$-electron wavefunction $\Psi(\underline{r},t)$ satisfies

$$i\hbar \frac{\partial \Psi}{\partial t} = H\Psi \quad \text{(II. 1)}$$

for Hamiltonian $H$ and $\underline{r} = \{\vec{r}_1, \vec{r}_2, ... \vec{r}_n\}$ is the electron configuration.

Because of the multiscale nature of a nanosystem, the wavefunction has multiple dependencies on $\underline{r}$ and time $t$. These dependences are made explicit via the hypothesis

$$\Psi(\underline{r},t) \to \Psi(\underline{r}, \underline{R}; t_0, \underline{t}; \varepsilon) \quad \text{(II. 2)}$$

Here, $\underline{R} = \varepsilon \underline{r}$ where $\varepsilon$ is the ratio of the shorter to longer characteristic lengths. The set of scaled times $\{t_0, \underline{t}\}$ (where $\underline{t} = \{t_1, t_2, ...\}$) is related to $t$ via $t_n = \varepsilon^n t$. The scaled particle configuration $\underline{R}$ is not an additional set of variables (i.e., there are not $6N$ degrees of freedom). Notable, the wavefunction $\Psi$ depends on $\underline{r}$ both directly and through $\underline{R}$ indirectly. Similarly, $\Psi$ depends on $t$ both directly and through the set of scaled times $\{t_0, \underline{t}\}$. The $\underline{r}$ dependence tracks the variation of $\Psi$ at the short-scale (here, typically on the scale of the average nearest-neighbor electron distance or the ion core lattice spacing). In contrast, the $\underline{R}$ dependence captures the long-range correlations. Similarly, $t_0$ tracks the shortest timescale, and the $t_n (n > 0)$ tracks the longer-scale ones.

Placing the multiscale ansatz Eq. (II. 2) into the Schrödinger equation Eq. (II. 1), one obtains an unfolded wave equation wherein $\varepsilon$ appears explicitly [9]. With this, $\Psi$ is constructed as a power series in $\varepsilon$. The theory has different implications, depending on whether ground state solutions to the lowest-order problem are unique or degenerate, as well as the way the potential is split into contributions of various orders in $\varepsilon$. Taking the ground state of the short-scale problem to be non-degenerate, and adopting the convention that its energy is zero, the lowest order wave function $\Psi_0$ has the form

$$\Psi_0 = \widehat{\Psi}(\underline{r})W(\underline{R},t). \quad \text{(II. 3)}$$

Here, $W$ plays the role of an envelope function modulating the short-scale dependence of $\widehat{\Psi}$.

The lowest order wavefunction $\Psi_0$ (Eq. (II. 3)) must satisfy antisymmetry with respect to particle exchange. There are three possible cases. (1) $\widehat{\Psi}$ is antisymmetric while $W$ is symmetric, (2) $\widehat{\Psi}$ is symmetric while $W$ is antisymmetric, and (3) both $\widehat{\Psi}$ and $W$ have mixed symmetry but their product can be antisymmetrized. Since plasmons are the focus of the present study, only case (1) is considered; this is found to imply that plasmons are bosons[31].

A first step in the multiscale analysis of a quantum nanosystem is to split the interaction forces into long and short range contributions. This splitting must be done in a manner that captures the phenomena of interest, and which facilitates the computer simulation of the otherwise challenging quantum nanosystems. Starting with a physical rational for the splitting and a specification of the type of phenomenon to be studied, the objective is to provide a rational for the scaling of space and time and thus the resulting CGWE.

Consider a two-body potential $v(r)$ that depends on interparticle distance $r$. Let $G(r)$ be a Gaussian-like function (i.e., $G(0)=1$ and $G \to 0$ as $r \to \infty$). Then $v$ is split via

$$v(r) = v(r)G(r) + v(r)[1-G(r)]. \quad \text{(II. 4)}$$

The first term has short-scale character, i.e., approaches zero as $r \to \infty$ faster than $v(r)$. However, the second term is well-behaved as $r \to 0$ assuming, by construction, that $1-G$ decays to zero as $r \to 0$ faster than $v(r) \to \infty$. By choice of the range over which $G$ decays, the second term can be considered a perturbation, i.e., contributes to $\varepsilon V_1$ rather than to $V_0$. For Coulomb systems such as electrons in a metal or semi-conductor, there is an additional reason why the $v(r)\left[1-G(r)\right]$ term can be considered a perturbation. As the distance between two electrons exceeds a few lattice spacing, the total potential (electron-electron plus electron-ion core) acts as a screened interaction. Thus, if one considers each electron to be near an oppositely charged ion core, then the net effect of the $v(r)\left[1-G(r)\right]$ terms summed over all the electrons and ion cores expresses much cancellation. As a result of this cancellation, it is hypothesized that the net result of those long-range interactions will bring a contribution that is a power of $\varepsilon$ smaller than that due to an individual pair interaction. This is the reason for choosing the following ansatz for the total potential $V$:

$$V = V_0 + \varepsilon^2 V_2. \quad \text{(II. 5)}$$

This ansatz has a strong implications for the $\varepsilon \to 0$ analysis as discussed earlier[9], and which will be justified quantitatively via the computational results presented in Sect. IV below.

Carrying the analysis to $O(\varepsilon^0)$, an equation for $\widehat{\Psi}$ emerges [9]:

$$i\hbar \frac{\partial \widehat{\Psi}}{\partial t_0} = H_0 \widehat{\Psi} \quad \text{(II. 6)}$$

where $H_0$ is short-scale Hamiltonian.

Carrying the analysis to $O(\varepsilon^2)$, an equation for $W$ emerges [9]:

$$\begin{aligned}
&i\hbar \frac{\partial W}{\partial t_2} = H^{CG} W, \quad H^{CG} = V^{CG} + \sum_{\ell,\ell'=1}^{N} \sum_{\alpha,\alpha'}^{3} \mu_{\ell\alpha\ell'\alpha'} \frac{\partial^2}{\partial R_{\ell\alpha} \partial R_{\ell'\alpha'}}, \quad V^{CG} = \langle 0|V_2|0\rangle, \\
&\mu_{\ell\alpha\ell'\alpha'} = -\frac{\hbar^2}{2m}\delta_{\ell\ell'}\delta_{\alpha\alpha'} + \widetilde{\chi}_{\alpha\alpha'}, \quad \widetilde{\chi}_{\alpha\alpha'} = \frac{i\hbar}{m^2} \int_{-\infty}^{0} dt_0 \chi_{\ell\alpha\ell'\alpha}(t_0), \quad \chi_{\ell\alpha\ell'\alpha'}(t_0) = \langle 0|p_{\ell\alpha} S(-t_0) p_{\ell'\alpha'}|0\rangle.
\end{aligned} \quad \text{(II. 7)}$$

In the above, $H^{CG}$ is the coarse-grained Hamiltonian, and $V^{CG}$ is the coarse-grained potential; $|0\rangle$ represents $\hat{\Psi}$ ground state, and $|n\rangle$ ($n>0$) represents an excited state of $H_0$, and $S(t_0)$ is the evolution operator $\exp(-i(H_0 - i0^+)t_0/\hbar)$. The positive infinitesimal $0^+$ is introduced to ensure causality. $p_{\ell\alpha}$ is the momentum of particle labeled $\ell$ along $\alpha$ direction. The $\mu_{\ell\alpha\ell'\alpha'}$ term corresponds to an effective inverse mass that is tensorial in character and introduces a two-body term (i.e., depends on $\ell$ and $\ell'$), and $m$ is the electron mass. The inverse effective mass tensor $\tilde{\chi}_{\alpha\alpha'}$ is [9]

$$\tilde{\chi}_{\alpha\alpha'} = \frac{\hbar^2}{m^2} \sum_{n \neq 0} \frac{\langle 0|p_\alpha|n\rangle\langle n|p_\alpha|0\rangle}{\zeta_n}. \qquad (\text{II. 8})$$

In the above equation

$$H_0|n\rangle = \zeta_n|n\rangle \qquad (\text{II. 9})$$

and the ground state energy $\zeta_0 = 0$.

In this two-scale deductive multiscale approach, the Hamiltonian of the system is partitioned into short-scale and long-scale contributions. The long-scale is coupled to the short-scale via the effective masses and interactions. The implementation of these results as a multiscale quantum simulator is discussed below.

### III. THE MULTISCALE COMPUTATIONAL ALGORITHM

The following algorithm based on the mathematical framework of Sect. II was developed and implemented using variational Monte Carlo method as follows (Fig. 1).

1. Construct the ion core lattice and short-range potential $V_0$.

2. Choose a short-scale trial function that includes Jastrow-like electron-electron and ion-electron correlation factors [30].

3. Use a variational Monte Carlo approach [32, 33] to optimize the ground and excited states of $H_0$, and determine associated energies.

4. Construct effective mass tensor and interactions.

5. Set up the long-scale simulation domain.

6. Construct trial functions for CGWE and optimize them via a Monte Carlo variational approach to obtain the ground and excited states of the CGWE.

7. Use the above to construct the excitation spectrum.

Details on the implementation are provided below.

**A. Ion core lattice and simulation domain for the short-scale problem**

The center of mass of the nanoparticle is positioned at the center of the simulation box. An empty zone (here, a thickness of two lattice spacings) is added to the rectangular simulation domain along each of the three Cartesian axis. It was found that increasing the empty zone further than two lattice spacings did not affect the accuracy of the ground state energy significantly, while it did introduce a greater computational burden.

As in Sect. II, the pair potential $v(r)$ was split into short- and long-range contributions as $v(r) = v_0(r) + \varepsilon^2 v_2(r)$. Using $G(r) = e^{-\kappa r^2}$ the short-range pair potential is

$$v_0(r) = v(r) e^{-\kappa r^2}, \qquad \text{(III. 1)}$$

while the long-range pair potential is given by

$$v_2(r) = v(r)(1-e^{-\kappa r^2})r/\varepsilon R \quad \text{(III. 2)}$$

where $R = \varepsilon r$. For ion-electron interaction a simple pseudopotntial was adopted[34, 35] via:

$$v^{ion}(r) = \begin{cases} 0, & r < r_c \\ -Ze/r, & r > r_c \end{cases}, \quad \text{(III. 3)}$$

where $r_c$ is the experimentally determined radius of an ion core and $Z$ is the valence charge. ($r_c = 1.09$ Å and $Z = 1$ for silver[36]).

The relative strength of short-range vs long-range interaction is determined by $\kappa$. If $\kappa$ is large, then the short-range problem for electrons in a finite ion core lattice does not support bound states. Alternatively, if $\kappa$ is small, then the short-scale problem reduces to the full Schrödinger equation and no benefit is attained via the multiscale approach. Similarly, the choice of $\kappa$ effects the bound state character of the long-scale problem. If $\kappa$ is too small, then $W$ is essentially the solution of a free-particle long-scale problem. Thus there are no bound states for $W$ and hence, again, the analysis does not directly apply to finite assemblies. One concludes that an intermediate value of $\kappa$ would give computational advantages for the short-scale problem and yield a viable theory for the coarse-grained wave function $W$ that preserves the character of the phenomena of interest, i.e., collective modes interacting with particle-like degrees of freedom. In the present demonstration, it was found that taking $\kappa^{-1/2}$ to be a few ion core lattice spacings was a good compromise. This choice of $\kappa$ yields bound states for the short- and long-scale problems for silver nanoparticles.

**B. Trial function for the short-scale problem**

The variational trial function adopted for the ground state of the short-scale problem was

$$\widehat{\Psi} = e^{-U(\underline{\eta},\underline{r})}S(\underline{r},\underline{\lambda}), \quad \text{(III. 4)}$$

where $S(\underline{r},\underline{\lambda})$ is a Slater determinant of single particle spin-orbitals and $U(\underline{\eta},\underline{r})$ is a symmetric function of the N-electron spin-configuration $\underline{r}$; $\underline{\eta}$ and $\underline{\lambda}$ are sets of variational parameters obtained by minimizing the energy $E(\underline{\eta},\underline{\lambda})$:

$$E(\underline{\eta},\underline{\lambda}) = \frac{\langle \widehat{\Psi} | H_0 | \widehat{\Psi} \rangle}{\langle \widehat{\Psi} | \widehat{\Psi} \rangle}. \qquad \text{(III. 5)}$$

The inner products $\langle \widehat{\Psi} | H_0 | \widehat{\Psi} \rangle$ and $\langle \widehat{\Psi} | \widehat{\Psi} \rangle$ were evaluated by Monte Carlo integration [32]. The form of the Jastrow-like factor $U(\underline{\eta},\underline{r})$ was

$$U(\underline{\eta},\underline{r}) = \eta_{\text{e-e}} \sum_{i \neq j}^{N_e} \frac{e^{-\kappa(r_i-r_j)^2}}{|r_i - r_j|} + \eta_{\text{ion-e}} \sum_{i,I}^{N_e,N_{\text{ion}}} \frac{e^{-\kappa(r_i-r_I)^2}}{|r_i - r_I|}, \qquad \text{(III. 6)}$$

where $\eta_{\text{e-e}}$ and $\eta_{\text{ion-e}}$ are parameters that capture electron-electron and ion-electron correlations, respectively. $r_i$ and $r_j$ are the position of electrons, and $r_I$ is the position of an ion. This form of $U(\underline{\eta},\underline{r})$ is similar to that adopted[37] for the polarized electron gas.

The Slater determinant is constructed from free-particle states for the box-shaped simulation domain. For example, the single particle state for electron $i$ was taken to be

$$\psi(x_i, y_i, z_i) = \sin(k_x x_i)\sin(k_y y_i)\sin(k_z z_i) \qquad \text{(III. 7)}$$

where $k_x = 2\pi n_x / L_x$ for integer $n_x$ (equals $1, 2, \cdots$) and width $L_x$ of the simulation box in the $x$ direction, and similarly for the $y$ and $z$ directions. The present study is restricted to closed-shell uncharged systems. Thus the number $N_e$ of electrons equals that of the ion cores $N_{\text{ion}}$, the latter assumed to have unit charge. Furthermore, $N = N_e = N_{\text{ion}}$ is considered to be even so that there are $N/2$ orbitals of spin up electrons and $N/2$ orbitals of spin down electrons. No additional variational parameters $\lambda$ were used for simplicity in this demonstration study.

## C. Variational Monte Carlo solution for the short-scale ground state

With the above trial function $\hat{\psi}$, the ground state total energy of the short-scale Hamiltonian of Eq. (II. 6) is computed by variational Monte Carlo method with a uniform sampling technique. The number of samples required to converge the Monte Carlo integration was determined as follows. Statistics was collected by repeating the Monte Carlo integration $M$ number of times. Each time the quantity $x_i = \langle |K.E| + |Vee| + |Vne| \rangle_n$ was evaluated with $n$ samples. Note that $x_i$ is not the total energy but the sum of absolute value of energy components. This was done to add the errors from each energy component. The coefficient of error $C = \sigma/\mu$ was computed, where $\sigma = (1/M)\sqrt{\sum_{i=1}^{M}(x_i - \mu)^2}$ was the standard deviation and $\mu = (1/M)\sum_{i=1}^{M} x_i$ was the mean value. The coefficient of variation $C$ was an indicator of percentage of error. In the present study, the Monte Carlo integration was performed to achieve an estimated error of about 3%. If the percentage of error was greater than 3%, the number of samples $n$ was increased for each $M$ integrations.

The variational parameters $\eta_{e-e}$ and $\eta_{ion-e}$ in the short-scale trial function was optimized to minimize the ground state energy. This was done by Nelder-Mead simplex [38] and iterative-bisection methods. The iterative-bisection is similar to Powell's derivative free multidimensional optimization[39] which performed line-minimization of one parameter at a time. The iterative bisection optimization was faster than Nelder-Mead simplex method if the bisection intervals were chosen close to the optimum. In practice, the simplex search was performed to get approximate values of $\eta_{e-e}$ and $\eta_{ion-e}$, then iterative bisection method proceeds to fine tune the optimal values.

## D. Excited states of the short-scale problem

Variational Monte Carlo can be used to construct accurate excited states of atomic and molecular systems when the trial functions satisfy orthogonality and orbital symmetry constraints[40, 41]. Thus, a trail function constructed from single-particle states and Jastrow-like factor was used as for $\widehat{\Psi}$ (Eq. (III. 6)), by replacing one of the single particle functions at the Fermi surface with one of wave vector corresponding to a state above the Fermi surface. Then the variational method is used to optimize these trial functions. This is repeated to get a set of excited states $\widehat{\Psi}_1, \widehat{\Psi}_2, \cdots$.

## E. Effective Masses and Interactions

The effective mass tensor that appears in the CGWE (Eqs. (II. 7) and (II. 8)) is constructed as in Sect. II. It was found that 20 excited states were sufficient to obtain a converged value of the effective mass tensor (Fig. 2). This is because the matrix elements $\langle \widehat{\Psi}_0 | \bar{p}_\ell | \widehat{\Psi}_N \rangle$ and $(E_n - E_0)^{-1}$ both decrease rapidly with $n$ (refer Eq. (II. 8)), a necessary condition for the viability of the DMA approach [9].

The effective interaction is computed via effective charges ($Q_{\text{eff}}^{\text{ion-e}}$ and $Q_{\text{eff}}^{\text{e-e}}$) as follows. As outlined in Sect. II. A the long-range ion-electron pair potential is written

$$v_2^{\text{ion-e}}(r, R) = v^{\text{ion-e}}(r)\left[1 - e^{-\kappa r^2}\right]\frac{r}{\varepsilon R}. \quad \text{(III. 8)}$$

For the pseudopotential considered here (Eq. (III. 3)),

$$v_2^{\text{ion-e}}(r, R) = \begin{cases} 0, & r < r_c \\ -\dfrac{\left[1 - e^{-\kappa r^2}\right]}{\varepsilon R}, & r > r_c \end{cases}. \quad \text{(III. 9)}$$

Construction of the coarse-grained pair potential is obtained by averaging $v_2^{\text{ion-e}}(r, R)$ with respect to $\widehat{\Psi}_0$ as given by

$$v^{\text{ion-e, CG}}(R) = -\frac{\left\langle \widehat{\Psi}_0 \left| \left[ 1 - e^{-\kappa r^2} \right] \right| \widehat{\Psi}_0 \right\rangle}{\varepsilon R} = -\frac{Q_{\text{eff}}^{\text{ion-e}}}{\varepsilon R}. \quad \text{(III. 10)}$$

Note that $v^{\text{ion-e, CG}}(R)$ is a function of the scaled coordinate $R$. The long range part of the e-e pair potential is given by

$$v_2^{\text{e-e}}(r, R) = v^{\text{e-e}}(r)\left[1 - e^{-\kappa r^2}\right]\frac{r}{\varepsilon R} = \frac{\left[1 - e^{-\kappa r^2}\right]}{\varepsilon R}. \quad \text{(III. 11)}$$

The coarse-grained pair potential is obtained by averaging $v_2^{\text{e-e}}(r, R)$ with respect to $\widehat{\Psi}_0$,

$$v^{\text{e-e, CG}}(R) = \frac{\left\langle \widehat{\Psi}_0 \left| \left[ 1 - e^{-\kappa r^2} \right] \right| \widehat{\Psi}_0 \right\rangle}{\varepsilon R} = \frac{Q_{\text{eff}}^{\text{e-e}}}{\varepsilon R}. \quad \text{(III. 12)}$$

The total coarse-grained potential $V^{CG}$ is computed from the sum $v^{\text{ion-e, CG}}(R)$ and $v^{\text{e-e, CG}}(R)$.

**F. Long-scale simulation domain**

The scaling parameter $\varepsilon$ is defined via

$$\varepsilon^2 = \frac{\kappa d^2}{1 + \kappa D^2} \quad \text{(III. 13)}$$

where $d$ is the short characteristic length of the system and, $D$ is its long characteristic length. The short characteristic length in a periodic lattice is the distance between the ion cores which for silver 2.88Å[36]. $D$ may be associated with mean free path, Fermi wavelength or extend of the electron density field beyond the metal surface. For silver these lengths are the mean free path ≈ 30 Å[42]; Fermi wavelength ≈ 5 Å[43]; and extend of the electron density field beyond the metal surface ≈5- 50 Å[43]. From the above long characteristic lengths, we believe the smallest should be used in order to capture all relevant longer scales i.e., the coarse-grained wave equation should express oscillations on these scales if they are captured by the approximations made and are relevant to the physical conditions. In the present study it was found that the electron density field extends about 15 Å beyond the silver nanoparticle surface. Thus, $D$ is chosen to be in the order of Fermi wavelength ($D \approx 2d$). The coarse-grained problem is cast in terms of the set of scaled electron positions $\underline{R} = \varepsilon \underline{r}$. Thus, the coarse-grained simulation domain is a rectangular box of edge lengths $\varepsilon L_x \times \varepsilon L_y \times \varepsilon L_z$. Thus, the coarse-grained simulation domain has volume $\varepsilon^3$ smaller than the short-scale one.

### G. Trial functions for the long-scale ground and excited states

As noted in Sect. II, the present development is for the case where the long-scale problem has bosonic character, i.e., the coarse-grained wave function W, has boson exchange symmetry. Adopting the coarse-grained mean field approximation [9], the trial function $W$ is a symmetrized product of single-particle functions. In particular, the coarse-grained ground state trail function $W_0$ is taken in the form

$$W_0 = \prod_{i=1}^{N} A(\vec{R_i}). \qquad \text{(III. 14)}$$

Similarly, the coarse-grianed excited state is taken in the form

$$W_1 = \prod_{i=1}^{N} A(\vec{R_i}) \sum_{j=1}^{N} B(\vec{R_j}) / A(\vec{R_j}). \qquad \text{(III. 15)}$$

Here, $A(\vec{R})$ is the ground state single-particle function and $B(\vec{R})$ is an excited sate single-particle function. With this, the trial functions $W_0$ and $W_1$ have bosonic exchange symmetry.

The following specific choices for the trial functions are made for spherical nanoparticles. Therefore the ground state function $W_0$ should have spherical symmetry. With this, the trial function for the single particle state is taken to be

$$A(\vec{R}) = [1 + \exp \nu (R - R_{max})]^{-1} \cos(\frac{\pi X}{\varepsilon L}) \cos(\frac{\pi Y}{\varepsilon L}) \cos(\frac{\pi Z}{\varepsilon L}) \qquad \text{(III. 16)}$$

for cubic simulation domain of edge length $\varepsilon L$, and where $R = |\vec{R}|$; $R_{max}$ and $\nu$ are variational parameters. The cosine factor ensures that $W_0$ vanishes at the domain boundary, while the $\nu$, $R_{max}$ factor reflects the spherical symmetry of the nanoparticle.

The excited state trial function is built using spherical harmonic functions, i.e.,

$$B_{\ell m}(\vec{R}) = A(\vec{R}) Y_\ell^m(\theta, \phi). \qquad \text{(III. 17)}$$

With this choice the excited states and the ground state are orthogonal, simplifying some of the computations.

## H. Variational solution of the long-scale problem

Having picked the form of single-particle functions (i.e., $A(\vec{R_i})$ and $B(\vec{R_j})$), the coarse-grained total energy is

$$E^{CG}(\nu, R_{max}) = \frac{\langle W | H^{CG} | W \rangle}{\langle W | W \rangle}. \qquad \text{(III. 18)}$$

The parameters $\nu$ and $R_{max}$ in the ground state trial function were optimized for the minimum energy. The procedure of optimization is similar to the procedure described in Sect. III. C via iterative bisection method. The optimization of parameter $R_{max}$ is greatly simplified by choosing appropriate lower and upper bounds. The choice of $R_{max}$ bound has been derived based on the characteristics of the Fermi function that appear in the ground state. The lower bound for $R_{max}$ is nearly equal to the radius of the nanoparticle and the upper bound is about one lattice distance away from the nanoparticle surface. Indeed, this choice of bound was found to be appropriate when iterative-bisection was performed with a large bisection interval and simplex optimization.

## IV. DEMONSTRATION FOR PLASMONS IN SILVER NANOPARTICLES

The multiscale algorithm of Sect. III was used to predict the plasmonics of spherical silver nanoparticles. The structure of minimum energy optimized silver nanoparticles with fewer than 100 atoms was obtained from the literature [44]. Structures with greater than 100 atoms were taken to be spherical fragments extracted from the bulk periodic lattice without geometry optimization. The latter was adopted since the internal structure of larger particles tends to resemble that of the bulk. For example, the geometry optimized silver cluster with 76 atoms resembles the bulk crystal structure [44] (Fig. 3). To reduce the magnitude of the variational calculations, each ion core was represented by a pseudopotential (Sect. III. A).

The value of $\kappa$ was chosen to ensure the existence of bound states for both the short- and long-scale Hamiltonians. In particular, $\kappa$ was chosen to be $1/(2d)^2$, where $d$ is the nearest neighbor distance between ion cores. For silver the lattice constant $a$ is 4.09 Å[36] and $d = a/\sqrt{2}$. The long characteristic length $D = 2d$, and therefore Eq. (III. 13) implies that $\varepsilon$ equals 0.125.

The short-scale problem was solved for systems with 8, 34, 64, and 76 silver atoms. For these systems, the ground state is non-degenerate so that the scaling approach of Sect. II applies. The ground and excited states for the short-scale Hamiltonian were constructed as outlined in Sect. III. B; the effective masses and interactions were constructed as in Sect. III. E. The variational parameters $\eta_{e-e}$ and $\eta_{ion-e}$ of the short-scale excited states were assumed to be same as that of ground state. For clusters with more than 64 atoms, the effective masses and interactions were found to be insensitive to cluster size. Thus, these quantities were evaluated for the 64-atom cluster and used to construct the CGWE for clusters containing 100 to 2106 atoms.

The coarse-grained ground and excited states obey boson-exchange symmetry, and were taken to be of coarse-grained mean field form[9](refer Eqs. (III. 14), (III. 15), (III. 16), and (III. 17)). The single particle excited state trial function includes a spherical harmonic factor $Y_\ell^m$ to characterize the plasmons in the spherical assembly (e.g., for dipolar, quadrupolar, and more complex excitations). For example, the dipolar mode $Y_1^0$ ($=Z/R$) involves a positive and a negative zone in the north and south hemispheres, respectively.

The dipolar plasmon excitation energy as a function of nanoparticle size as predicted by the multiscale computation is shown in Fig. 4. Also in the figure, similar values for TDDFT calaculations [45] and experiments [46, 47] are shown for comparison. In all cases, the plasmon energy decreases with nanoparticle size. The excitation energies predicted by the present and TDDFT[45] approaches are within an order of magnitude of the observations.

Discrepancies between predictions and observations likely come from several sources. The experimental conditions such as temperature, solvents, and impurities are not accounted for in the theory. The structure of the nanoparticle under experimental investigation could be different from that used in the theoretical studies. Moreover, the psueopotential used neglects effects from core d electrons[48]. Simple forms of short- and long-scale trial functions were used. Since the multiscale approach was successful in predicting the correct trend and order of magnitude of the plasmon spectrum, the quantum multiscale approach appears to be a practical method for simulating plasmon excitations for a broad range of nanoparticle sizes. Results will likely improve when the above possible sources of error are addressed.

## V. CONCLUSIONS

It was shown that multiscale techniques deduced from the Schrödinger equation could provide quantitative predictions of the large scale, low-lying excitations of a delocalized, nanoscale electronic system. The coupling of individual particle (i.e., short-scale) processes with long-scale ones was accounted for via the coupling of a short fermionic wave equation with a long-scale bosonic one. Order of magnitude agreement for the plasmon spectrum of silver nanoparticles was achieved with a caliberation-free calculation. In this approach, plasmons emerge as bosonic excitations of the aforementioned long-scale problem. This multiscale approach is applicable to arbitrary nanoparticle shape and composition. This makes the theory suitable for applications to nanoscale electronic, plasmonic, superconducting, and energy storage materials. Recent multiscale theoretical results similar to that on which the present work is based[11] open the way to the analysis of BCS superconductors by accounting for the coupling of long-scale electronic and slow nuclear motions. To make the theoretical computational approach applicable to some of the above systems at finite temperature, the theory must be generalized via density matrix formalism.

The scaling of quantum Monte Carlo (QMC) computation with system size ($N$) is approximately $N^3$ [49]. This scaling is due to the effort involved in evaluating the Slater determinant. The largest system investigated with the QMC approach is 1,000 electrons [50]. The quantum multiscale approach reduces the computation effort by dividing the full problem into a short-scale fermion and a long-scale boson calculation. Although solving the short-scale problem is an $N^3$ effort, the long-scale one is an $N^2$ effort since it does not involve computing the Slater determinant. The $N^2$ scaling by the bosonic computation comes from the two-body kinetic and potential energy terms. While the short-scale problem scales as $N^3$, the effective masses and interactions derived from it rapidly become independent of $N$. Thus, the overall computation is $N^2$ and not $N^3$. An additional factor in accessing the efficiency of the present approach is that the bosonic long-scale problem does not need the high sampling density required by direct Monte Carlo.

Further advances with the present multiscale computational approach include the following:

- Improved trial functions for the long and short scale problems.

- Accounting for more complex families of excitations wherein the long- and short-scale solutions have mixed exchange symmetry [9].

- Investigating cases where the ground state of the short-scale problem is degenerate with resulting Dirac equation-like behavior of the long-scale problem[9, 10].

- Simulating electron-phonon processes such as in BCS superconductors.

With this, we suggest that the present methodology should be of great interest in pure and applied studies.

**ACKNOWLEDGEMENTS**
This work was supported in part by grant number CHE 1037383 from the Theory, Models and Computational Methods Program of the National Science Foundation and the College of Arts and Sciences of Indiana University, Bloomington.

**FIGURES**

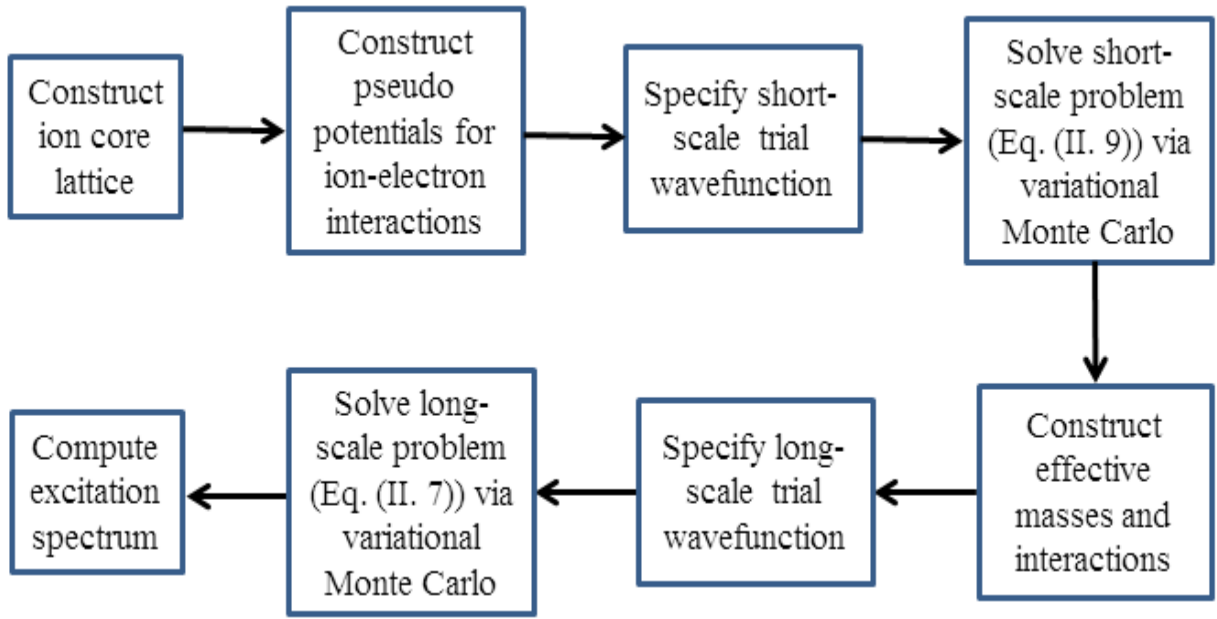

**Figure 1** Flow chart for the multiscale computational algorithm based on Sects. II and III.

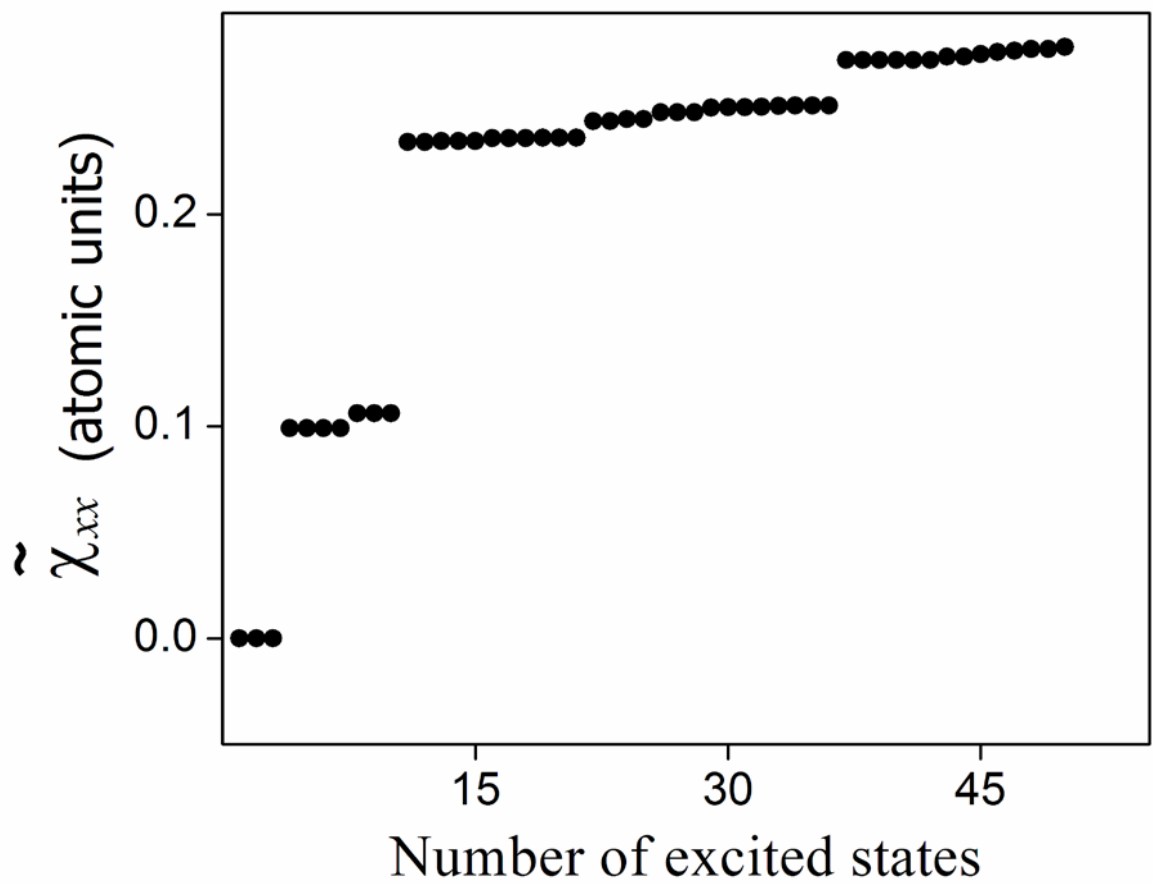

Figure 2 Convergence of inverse effective mass tensor component $\tilde{\chi}_{xx}$ for 76-atom silver cluster as a function of the number of excited states.

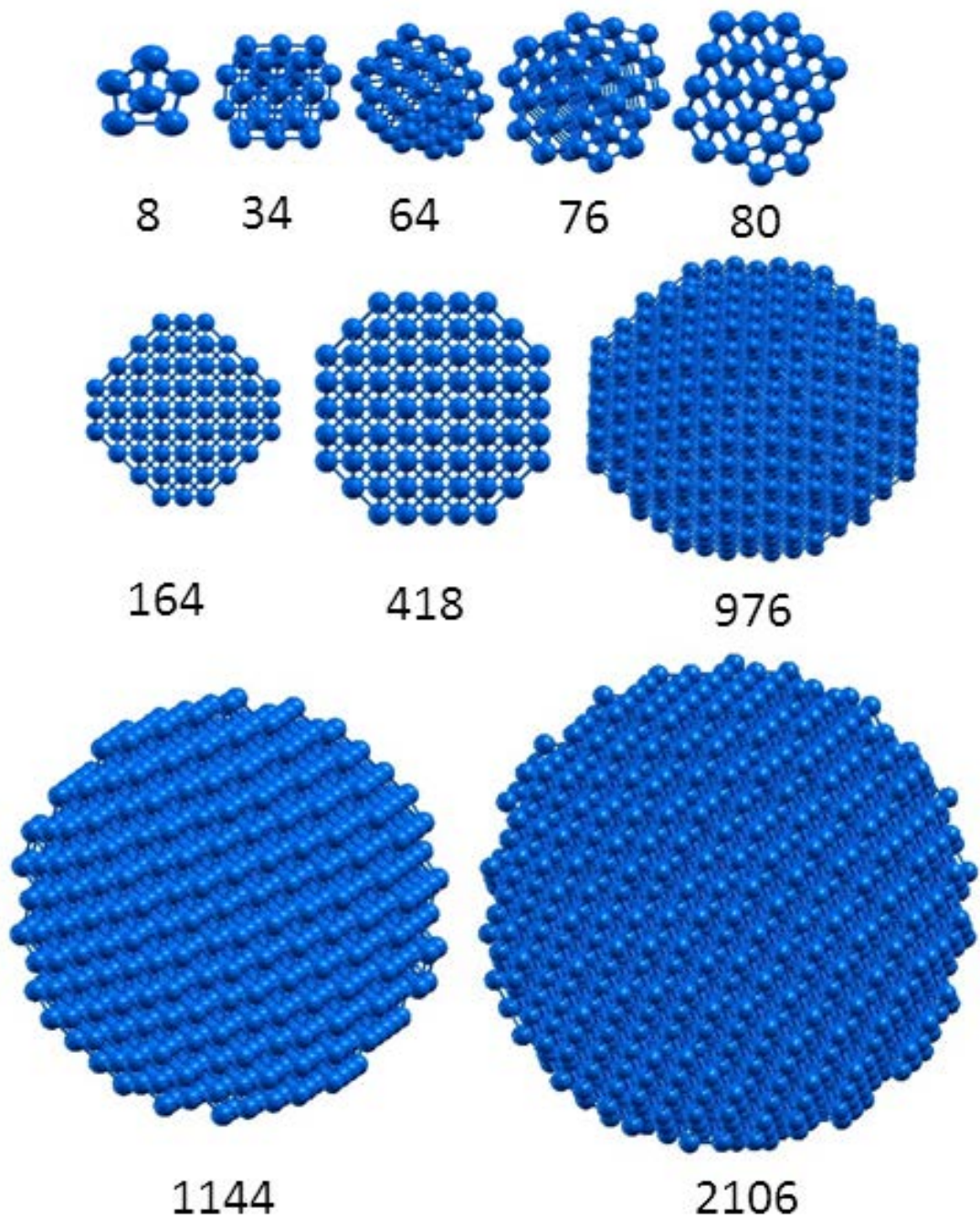

**Figure 3 The structure of silver nanoparticles. The number of atoms in a nanoparticle is given below each structure.**

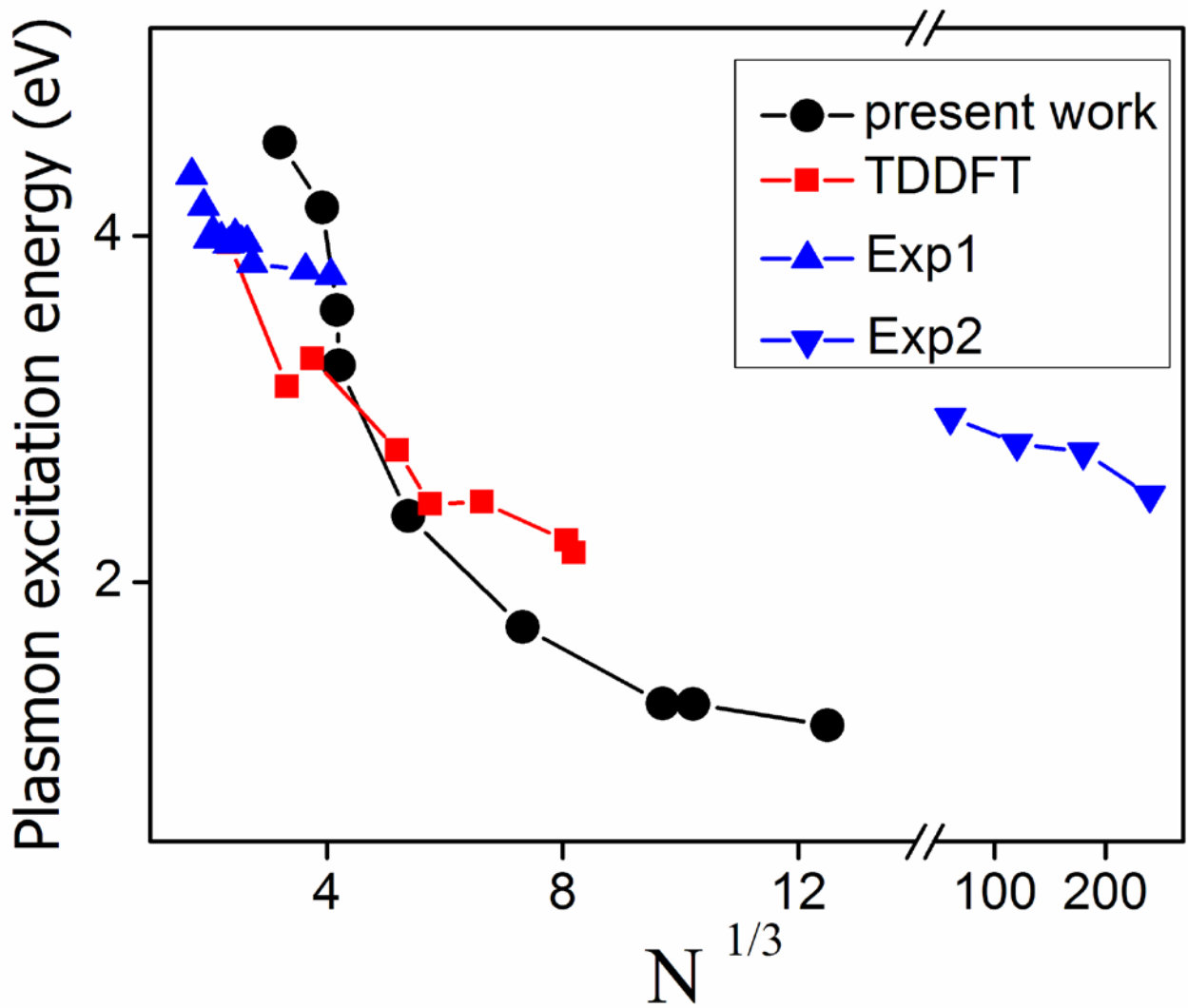

**Figure 4** Size dependence of the plasmon excitation energy for the dipolar mode in spherical silver nanoparticles. N is the number of silver atoms in a nanoparticle. Our predictions (circle) are compared with data from TDDFT [45] (square) and experiments (upward[47] and downward[46] pointing triangles).